\documentclass{article}

\input pz.sty
\input epsf.sty
\usepackage{longtable}
\font\fn=cmr12

\begin{document}

\PZhead{9}{27}{2008}{22 October}

\PZtitletl{Photometric investigation of bright}{type II-P 
Supernova 2004\lowercase{dj}}

\PZauth{D. Yu. Tsvetkov, V. P. Goranskiy, N. N. Pavlyuk}
\PZinsto{Sternberg Astronomical Institute, University Ave. 13,
119992 Moscow, Russia}

\begin{abstract}
CCD $UBVRI$ photometry is presented for type II SN 2004dj for about 
1200 days, starting day 2 past discovery. The photometric behaviour
is typical for SNe II-P, although some minor peculiarities are 
noticed. We compare the photometric data for the host cluster S96
before and after SN 2004dj outburst and do not find any significant
changes. 

\end{abstract}
\bigskip
\bigskip
\PZsubtitle{Introduction}

The brightest supernova of the past decade SN 2004dj was discovered by
K.Itagaki (Nakano, 2004) on 2004 July 31.76 UT in the nearby
SBcd galaxy NGC 2403. The spectra taken immediately after discovery
indicated it to be type II-P event found long after the outburst 
(Patat et al., 2004). The object was also detected in radio
(Stockdale et al., 2004), infrared (Sugerman, Van Dyk, 2005; Kotak 
et al., 2005) and X-ray bands (Pooley, Lewin, 2004). 
The optical photometry for SN 2004dj was published by Korc\'akov\'a et al.
(2005), Zhang et al. (2006), Vink\'o et al. (2006). Spectroscopic observations
were reported by Vink\'o et al. (2006) and Korc\'akov\'a et al. (2005).
The results show that SN 2004dj is a typical SN II-P, regarding both 
photometric and spectral evolution. The ejected mass is estimated to 
be about 10 $M_{\odot}$, and the mass of synthesized $^{56}$Ni 
$\sim 0.02 M_{\odot}$.  
$BVR$ light
curves and spectra in the nebular phase were presented by Chugai et al.
(2005), they pointed out the strong asymmetry of the H$\alpha$ emission line
at the nebular epoch.   
The photometric observations by Chugai et al. (2005) were reprocessed by us,
and the magnitudes presented here supersede the data reported in
Chugai et al. (2005). 
Spectropolarimetry reported
by Leonard et al. (2006) indicates strong departure from spherical
symmetry for the inner ejecta. Asymmetry of $^{56}$Ni ejecta that results
in the observed asymmetry of the H$\alpha$ emission line and the possibility
that this effect can also account for the polarization of SN radiation was
discussed by Chugai (2006).  

The association of SN 2004dj with the compact cluster Sandage 96 attracted
particular attention, the data on this cluster were reported by
Yamaoka et al. (2004), Ma\'iz-Apellan\'iz et al. (2004), Wang et al. (2005),
Chugai et al. (2005), and Vink\'o et al. (2006). The data suggests 
a cluster age of 14 - 20 Myr,
which results in probable SN progenitor mass of 12 - 15 $M_{\odot}$.   

\bigskip
\bigskip
\PZsubtitle{Observations and reductions} 

We started observations of SN 2004dj on 2004 August 2, two days after 
the discovery, but the field was also imaged at 1-m reflector of SAO
on 2001 January 19, long before the explosion.  

The observations of supernova were carried out with the 
following telescopes and CCD
cameras: 1-meter reflector of Special Astrophysical Observatory equipped
with CCD EEV42-40 (S100) (only the images obtained before SN discovery 
were made with CCD Electronika K-585); 70-cm reflector of SAI in Moscow (M70)
with Apogee AP-7p (a) or AP-47p (b) camera; 60-cm reflector of Crimean
Observatory of SAI (C60) with Princeton Instruments VersArrayB1300 (c), 
AP-47p , AP-7p,  or
SBIG ST-7 (d) CCD cameras; 50-cm Maksutov telescope of Crimean observatory
of SAI with Meade Pictor 416XT CCD camera (C50).

During three years of observations different filter sets were used at M70 and
C60, 
they are identified by numbers after the code for telescope
and CCD camera.
The color terms were derived by solving equations presented 
by Tsvetkov et al. (2006), they are reported in Table 1.
The observations at C50
were carried out only with $V$ filter which 
was close to standard system, and no correction was applied.

{\renewcommand{\arraystretch}{0}
\begin{table}
\centering
\caption{Color terms for different telescope/camera/
filter combinations}\vskip2mm
\fn
\begin{tabular}{lccr@{.}lr@{.}lc}
\hline
\rule{0pt}{4pt}\\
Code  &$K_u$& $K_b$ & \multicolumn{2}{c}{$K_v$} & 
\multicolumn{2}{c}{$K_r$} & $K_i$ \\
\rule{0pt}{4pt} \\ 
\hline
\rule{0pt}{4pt} \\ 
\strut M70a1 &      &  -0.11&  -0&032&  -0&24  & -0.38 \\ 
\strut M70a2 &      &  -0.14&  -0&023&  -0&12  & -0.38 \\
\strut M70b  & -0.05&  -0.21&  -0&023&   0&09  & -0.39 \\  
\strut C60a  &      &  -0.10&  -0&002&  -0&45  & -0.37 \\
\strut C60b1 & -0.03&  -0.21&  -0&017&   0&09  & -0.38 \\
\strut C60b2 & -0.05&  -0.16&  -0&033&  -0&03  & -0.45 \\
\strut C60c  & -0.06&  -0.10&  -0&026&  -0&025 & -0.43 \\
\strut C60d  &      &  -0.34&  -0&02 &   0&05  &       \\ 
\strut S100  &      &  -0.10&   0&09 &   0&036 &       \\
\hline
\end{tabular}
\end{table}
}
The standard image reductions and photometry were made using IRAF.\PZfm 
\PZfoot{IRAF is distributed by the National Optical Astronomy Observatory,
which is operated by AURA under cooperative agreement with the
National Science Foundation}

Photometric measurements of SNe were made relative to 
local standard stars using PSF-fitting with IRAF DAOPHOT package. 
We did not try to subtract the prediscovery images from the
images with supernova.

The magnitudes of local standard stars were calibrated on photometric
nights, when photometric standards were observed at different 
airmasses. They are presented in Table~2, the image of SN with 
marked local standards is shown in Fig~1.         

\PZfig{12cm}{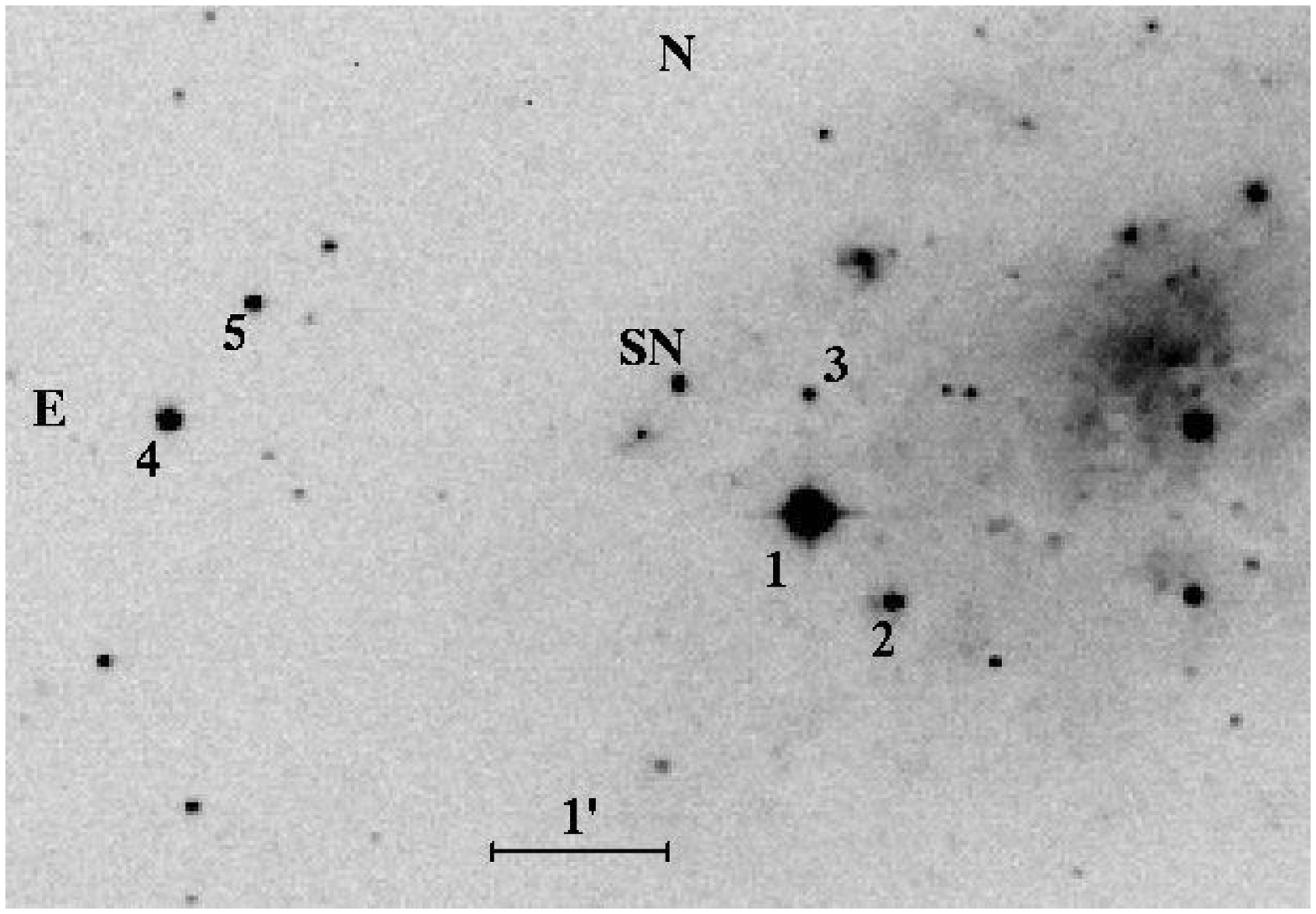}{SN 2004dj with local standard
stars}

{\renewcommand{\arraystretch}{0}
\begin{table}
\centering
\caption{Magnitudes of local standard stars}\vskip2mm
\fn
\begin{tabular}{cccccrcrcrc}
\hline
\rule{0pt}{4pt}&&&&&&&&&& \\
Star & $U$ & $\sigma_U$& $B$ & $\sigma_B$ &  \multicolumn{1}{c}{$V$}& 
$\sigma_V$ 
& \multicolumn{1}{c}{$R$} & $\sigma_R$ & \multicolumn{1}{c}{$I$} & $\sigma_I$\\
\rule{0pt}{4pt}&&&&&&&&&& \\
\hline
\rule{0pt}{4pt}&&&&&&&&&& \\
\strut 1 &   10.43 &0.02&  10.48& 0.02 &  9.95& 0.02 &  9.62& 0.02 &  9.35& 0.02\\
\strut 2 &   14.82 &0.07&  14.72& 0.03 & 14.08& 0.02 & 13.68& 0.02 & 13.35& 0.02\\
\strut 3 &   17.21 &0.13&  16.48& 0.03 & 15.56& 0.02 & 15.05& 0.02 & 14.61& 0.02\\
\strut 4 &   14.78 &0.02&  14.31& 0.02 & 13.50& 0.01 & 12.97& 0.04 & 12.63& 0.02\\
\strut 5 &   15.78 &0.04&  15.45& 0.02 & 14.75& 0.02 & 14.28& 0.04 & 13.98& 0.02\\
\hline
\end{tabular}
\end{table}
}

The magnitudes for our stars 1, 2, 3, 5 were 
derived by Vink\'o et al. (2006), and for stars 2, 3, 4, 5  by Stetson.\PZfm
\PZfoot{http://www2.cadc-ccda.hia-iha.nrc-cnrc.gc.ca/
community/STETSON/standards/}
The differences between our magnitudes and those from Vink\'o et al. (2006)
are quite significant, especially in the $B$ band, the mean differences are:
$\overline{\Delta B}= -0.14 \pm 0.03; \overline{\Delta V}= -0.09 \pm 0.02;
\overline{\Delta R}=-0.05 \pm 0.02; \overline{\Delta I}=-0.06 \pm 0.01$. 
The magnitudes from Stetson are in much better agreement with our data,
the mean differences are: $\overline{\Delta B}= -0.02 \pm 0.01;
\overline{\Delta V}= 0.00 \pm 0.01;
\overline{\Delta R}=0.01 \pm 0.02$.

The good agreement of our data with the magnitudes from Stetson 
suggests that our calibration is more reliable
than that by Vink\'o et al. (2006).

The photometry for SN 2004dj is reported in Table 3.  
 
\begin{center}
\begin{longtable}{cccccccccccl}
\caption{Observations of SN 2004dj}\\
\hline
JD 2450000+ & $U$ & $\sigma_U$ &
$B$ & $\sigma_B$ & $V$ & $\sigma_V$ & $R$ & $\sigma_R$ &
$I$ & $\sigma_I$ & Tel.\\
\hline
\endfirsthead

\multicolumn{12}{c}{\tablename\ \thetable{} -- continued from previous page}\\
\hline
JD 2450000+ & $U$ & $\sigma_U$ &
$B$ & $\sigma_B$ & $V$ & $\sigma_V$ & $R$ & $\sigma_R$ &
$I$ & $\sigma_I$ & Tel.\\
\hline
\endhead

\hline
\endfoot

\hline
\endlastfoot

1929.48 &      &     &  18.61& 0.04&  18.22& 0.03&  17.82& 0.03&       &     &  S100 \\
3220.31 &      &     &  12.49& 0.02&  11.89& 0.02&  11.51& 0.02&  11.36& 0.02&  M70a1 \\
3222.32 &      &     &  12.49& 0.02&  11.89& 0.02&  11.51& 0.02&  11.33& 0.02&  M70a1 \\
3238.27 &      &     &  12.92& 0.03&  11.96& 0.02&  11.54& 0.02&  11.34& 0.02&  M70a1 \\
3242.52 &      &     &  12.90& 0.02&  11.97& 0.02&  11.58& 0.02&  11.31& 0.02&  M70a1 \\
3244.57 & 13.80& 0.03&  12.91& 0.02&  11.97& 0.02&  11.53& 0.02&       &     &  C60c \\
3248.55 & 13.92& 0.06&  12.95& 0.02&  12.01& 0.02&  11.61& 0.02&       &     &  C60c \\
3249.47 &      &     &  12.97& 0.02&  12.05& 0.02&  11.59& 0.02&  11.44& 0.02&  M70a1 \\
3250.48 &      &     &       &     &  12.03& 0.02&       &     &       &     &  C50  \\
3251.51 &      &     &       &     &  12.06& 0.02&       &     &       &     &  C50  \\
3253.47 &      &     &       &     &  12.05& 0.02&       &     &       &     &  C50  \\
3254.52 &      &     &  13.02& 0.02&  12.07& 0.02&  11.65& 0.02&  11.36& 0.02&  M70a1 \\
3255.50 &      &     &       &     &  12.07& 0.02&       &     &       &     &  C50  \\
3255.56 & 14.11& 0.05&  13.09& 0.03&  12.07& 0.02&  11.69& 0.02&  11.40& 0.02&  C60c \\
3256.24 &      &     &       &     &  12.10& 0.02&       &     &       &     &  C50  \\
3257.22 &      &     &       &     &  12.09& 0.03&       &     &       &     &  C50  \\ 
3257.58 & 14.20& 0.04&  13.11& 0.03&  12.11& 0.02&  11.65& 0.02&  11.41& 0.03&  C60c \\ 
3259.60 &      &     &       &     &  12.12& 0.02&  11.71& 0.03&  11.46& 0.03&  C60c \\ 
3261.60 & 14.20& 0.11&  13.12& 0.03&  12.15& 0.02&  11.70& 0.03&  11.41& 0.03&  C60c \\ 
3263.54 &      &     &  13.20& 0.02&  12.22& 0.02&  11.79& 0.02&  11.56& 0.02&  M70a1 \\
3269.51 &      &     &  13.32& 0.03&  12.32& 0.02&  11.89& 0.02&  11.60& 0.02&  M70a1 \\
3294.49 & 16.16& 0.09&  14.75& 0.03&  13.79& 0.03&  13.36& 0.02&  13.01& 0.03&  M70b \\ 
3307.49 &      &     &  15.98& 0.03&  14.63& 0.02&  14.21& 0.02&  13.78& 0.03&  M70b \\ 
3309.63 &      &     &  15.95& 0.03&  14.67& 0.02&  14.16& 0.05&  13.74& 0.02&  C60c \\ 
3312.60 &      &     &  15.94& 0.03&  14.72& 0.02&  14.23& 0.02&  13.72& 0.03&  M70a1 \\
3315.57 & 17.16& 0.10&  15.96& 0.03&  14.74& 0.02&  14.22& 0.02&  13.80& 0.03&  C60c \\ 
3317.51 &      &     &  15.98& 0.03&  14.81& 0.02&  14.29& 0.02&  13.86& 0.03&  C60a \\ 
3318.54 & 17.11& 0.15&  15.98& 0.03&  14.77& 0.02&  14.34& 0.02&  13.82& 0.03&  C60c \\ 
3320.52 & 17.31& 0.09&  16.02& 0.03&  14.80& 0.02&  14.31& 0.02&  13.83& 0.02&  C60c \\ 
3321.56 & 17.32& 0.13&  16.03& 0.03&  14.81& 0.03&  14.33& 0.04&  13.85& 0.03&  C60c \\ 
3321.62 &      &     &  16.07& 0.02&  14.84& 0.02&  14.27& 0.02&       &     &  S100 \\ 
3323.53 & 17.30& 0.09&  16.03& 0.04&  14.85& 0.03&  14.38& 0.03&  13.85& 0.03&  C60c \\ 
3331.53 & 17.35& 0.10&  16.09& 0.03&  14.93& 0.02&  14.39& 0.02&  13.86& 0.02&  C60c \\ 
3355.54 &      &     &  16.27& 0.03&  15.21& 0.03&  14.45& 0.02&       &     &  C60c \\ 
3357.54 &      &     &  16.29& 0.02&  15.21& 0.02&  14.44& 0.03&       &     &  C60c \\ 
3358.54 &      &     &  16.26& 0.02&  15.19& 0.03&  14.45& 0.02&       &     &  C60c \\ 
3361.57 &      &     &  16.30& 0.04&  15.23& 0.03&  14.52& 0.03&       &     &  C60c \\ 
3385.36 &      &     &  16.53& 0.02&  15.46& 0.02&  14.53& 0.02&       &     &  S100 \\ 
3386.38 &      &     &  16.48& 0.02&  15.44& 0.02&  14.51& 0.02&       &     &  S100 \\ 
3387.42 &      &     &  16.52& 0.02&  15.45& 0.02&  14.51& 0.02&       &     &  S100 \\ 
3389.40 &      &     &  16.41& 0.04&  15.49& 0.03&  14.52& 0.02&  14.09& 0.02&  M70b \\ 
3406.33 &      &     &  16.46& 0.03&  15.56& 0.03&  14.59& 0.02&  14.15& 0.02&  M70b \\ 
3412.28 &      &     &  16.45& 0.05&  15.67& 0.04&  14.61& 0.03&  14.16& 0.03&  M70b \\ 
3427.38 &      &     &  16.58& 0.06&  15.68& 0.05&  14.67& 0.04&  14.21& 0.03&  M70b \\ 
3432.38 &      &     &  16.50& 0.08&  15.70& 0.05&  14.72& 0.04&  14.31& 0.04&  M70b \\ 
3436.49 &      &     &  16.63& 0.02&  15.75& 0.02&  14.83& 0.02&       &     &  S100 \\ 
3437.27 &      &     &  16.62& 0.02&  15.77& 0.02&  14.86& 0.02&       &     &  S100 \\ 
3444.28 &      &     &  16.72& 0.06&  15.83& 0.05&  14.80& 0.04&  14.41& 0.03&  M70b \\ 
3446.40 &      &     &  16.71& 0.04&  15.87& 0.04&  14.82& 0.03&  14.43& 0.03&  M70b \\ 
3456.40 &      &     &  16.67& 0.04&  15.95& 0.04&  14.87& 0.03&  14.46& 0.03&  M70b \\ 
3465.33 &      &     &  16.71& 0.05&  16.02& 0.06&  14.98& 0.05&  14.62& 0.06&  M70b \\ 
3500.40 &      &     &       &     &  16.21& 0.03&       &     &       &     &  S100 \\ 
3509.32 &      &     &  16.92& 0.05&  16.22& 0.06&  15.35& 0.04&  14.97& 0.05&  M70a2 \\
3530.27 &      &     &  17.03& 0.03&  16.33& 0.03&  15.51& 0.03&       &     &  S100 \\ 
3580.46 &      &     &       &     &  16.75& 0.07&  15.92& 0.04&  15.71& 0.05&  M70a2 \\
3581.53 &      &     &       &     &  16.75& 0.03&       &     &       &     &  C50  \\ 
3587.47 &      &     &  17.31& 0.05&  16.75& 0.04&  15.96& 0.04&  15.75& 0.04&  M70a2 \\
3588.54 &      &     &  17.34& 0.03&  16.84& 0.02&  16.03& 0.02&       &     &  C60b1 \\
3593.46 &      &     &  17.34& 0.04&  16.77& 0.07&  15.97& 0.03&  15.87& 0.05&  M70a2 \\
3593.54 &      &     &  17.34& 0.03&  16.96& 0.03&  16.10& 0.02&       &     &  C60b1 \\
3597.54 &      &     &  17.52& 0.04&  16.98& 0.02&  16.12& 0.03&       &     &  C60b1 \\
3607.52 &      &     &  17.27& 0.05&  16.98& 0.06&  16.06& 0.04&  15.89& 0.07&  M70a2 \\
3612.55 &      &     &  17.49& 0.04&  16.99& 0.04&  16.26& 0.04&  16.06& 0.07&  C60b1 \\
3621.44 &      &     &  17.39& 0.09&  17.02& 0.08&  16.20& 0.06&  16.03& 0.06&  M70a2 \\
3621.58 &      &     &       &     &       &     &  16.35& 0.06&       &     &  C60c \\ 
3628.50 &      &     &  17.35& 0.04&  17.08& 0.04&  16.26& 0.03&  16.06& 0.04&  M70a2 \\
3647.48 &      &     &  17.55& 0.05&  17.15& 0.06&  16.29& 0.05&  16.13& 0.08&  M70a2 \\
3676.55 &      &     &  17.67& 0.04&  17.40& 0.03&  16.70& 0.03&  16.47& 0.05&  C60b1 \\
3684.44 &      &     &  17.80& 0.04&  17.46& 0.04&  16.76& 0.03&  16.54& 0.04&  C60b1 \\
3728.44 &      &     &  17.88& 0.03&  17.63& 0.03&  17.05& 0.02&       &     &  C60c \\ 
3737.37 &      &     &       &     &  17.65& 0.40&  17.16& 0.07&       &     &  C60d \\ 
3738.43 &      &     &  18.00& 0.06&  17.66& 0.05&  17.10& 0.03&       &     &  C60d \\ 
3744.42 &      &     &       &     &       &     &  17.09& 0.08&       &     &  C60d \\ 
3801.47 &      &     &  17.95& 0.10&  18.09& 0.15&  17.30& 0.09&  16.95& 0.10&  M70a2 \\
3822.27 &      &     &  18.12& 0.08&  17.84& 0.10&  17.24& 0.06&  17.00& 0.08&  M70a2 \\
3826.30 &      &     &  17.90& 0.03&  17.72& 0.03&  17.27& 0.03&       &     &  S100 \\ 
3827.26 &      &     &  18.13& 0.04&  17.84& 0.03&  17.40& 0.03&       &     &  S100 \\ 
3859.33 &      &     &  18.29& 0.16&  17.94& 0.15&  17.36& 0.10&  16.96& 0.07&  M70a2 \\
3886.27 &      &     &  18.07& 0.03&  17.82& 0.02&  17.42& 0.02&       &     &  S100 \\ 
3887.27 &      &     &  18.12& 0.04&  17.93& 0.03&  17.45& 0.03&       &     &  S100 \\ 
3972.54 &      &     &       &     &  17.96& 0.03&       &     &       &     &  C50  \\ 
4034.76 &      &     &  18.22& 0.04&  17.97& 0.03&  17.60& 0.03&       &     &  S100 \\ 
4035.76 &      &     &  18.32& 0.04&  18.07& 0.03&  17.74& 0.03&       &     &  S100 \\ 
4086.54 &      &     &  18.03& 0.07&  17.97& 0.04&  17.53& 0.04&       &     &  C60c \\ 
4091.57 &      &     &  18.15& 0.04&  17.87& 0.04&  17.52& 0.05&       &     &  C60c \\ 
4146.47 &      &     &  18.36& 0.03&  18.09& 0.03&  17.72& 0.02&       &     &  S100 \\ 
4181.44 &      &     &  18.40& 0.03&  18.09& 0.03&  17.76& 0.03&       &     &  S100 \\ 
4418.52 &      &     &  18.31& 0.04&  18.12& 0.03&  17.77& 0.03&  17.32& 0.11&  C60b2 \\
4426.46 &      &     &       &     &  18.00& 0.04&       &     &       &     &  C50  \\ 
4428.54 &      &     &       &     &  18.03& 0.05&       &     &       &     &  C50  \\ 
\end{longtable}
\end{center}  

\bigskip
\bigskip
\PZsubtitle{The light and color curves}

The light curves are presented in Fig. 2. They are typical for SNe II-P, 
but only a small part of the plateau was covered by
observations. After the fast decline from the plateau the prominent
flattening, or secondary plateau, is evident on the light curves in
the $R$ and $I$ bands, which lasts about 160 days, and only after about 
JD 2453480 the linear decline begins.
At about JD 2453800 the light curves in all bands flatten, as the 
cluster S96 becomes the dominant source of luminosity.
We can subtract the luminosity of the cluster
from magnitudes obtained for the sum of cluster and supernova.
We used for subtraction the $B,V,R$ magnitudes of S96 derived from
images obtained before the explosion, and we adopted $I_{S96}=17.32$ from
our last image in this band.
Resulting light curves are shown in Fig. 3. The linear fits to the 
magnitudes in the period JD 2453500-800 give the following decline rates
(in mag day$^{-1}$):
0.0063 in the $B$, 0.0096 in the $V$,
0.010 in the $R$ and 0.011 in the $I$ band.   
In all bands except $B$ the rate is very close to the decay slope
of $^{56}$Co, which is 0.0098 mag day$^{-1}$.

\PZfig{12cm}{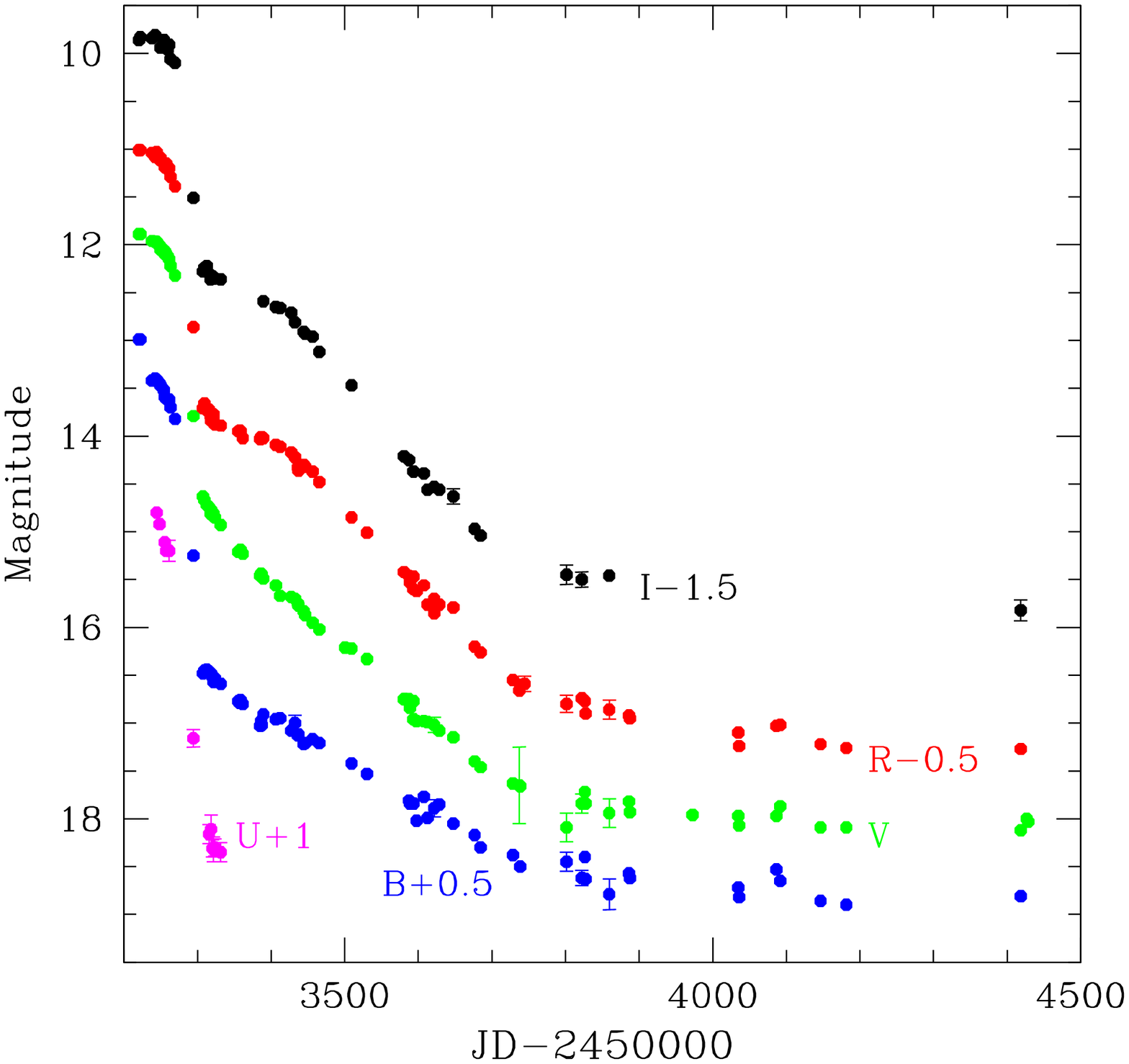}{$UBVRI$ light curves for SN 2004dj. The error bars
are shown only if they exceed the size of a point on this and the following 
figures}

\PZfig{12cm}{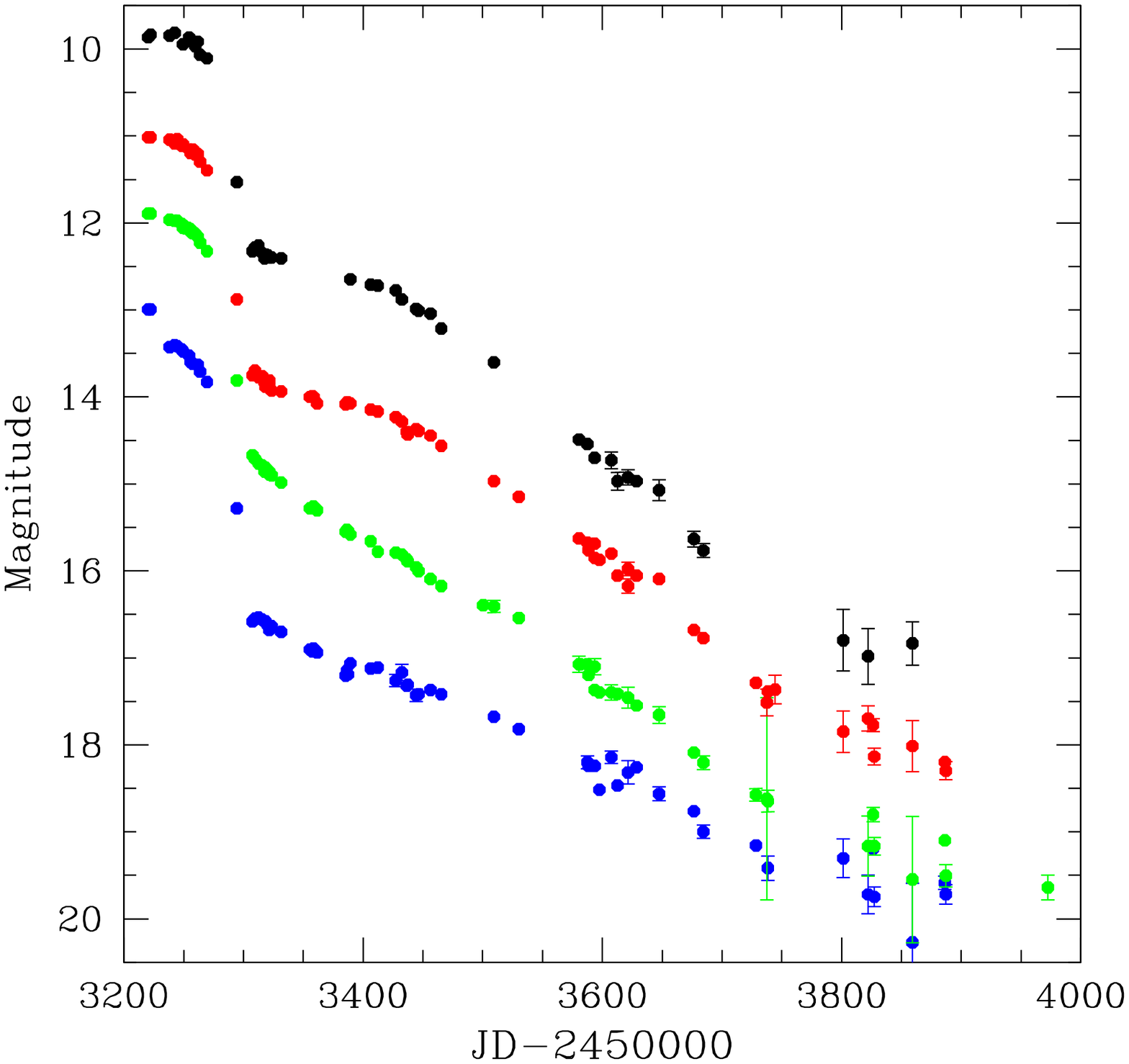}{$BVRI$ light curves for SN 2004dj after subtraction
of the luminosity of the cluster S96. The color coding and shifts of magnitudes
are the same as on Fig. 2}

\PZfig{12cm}{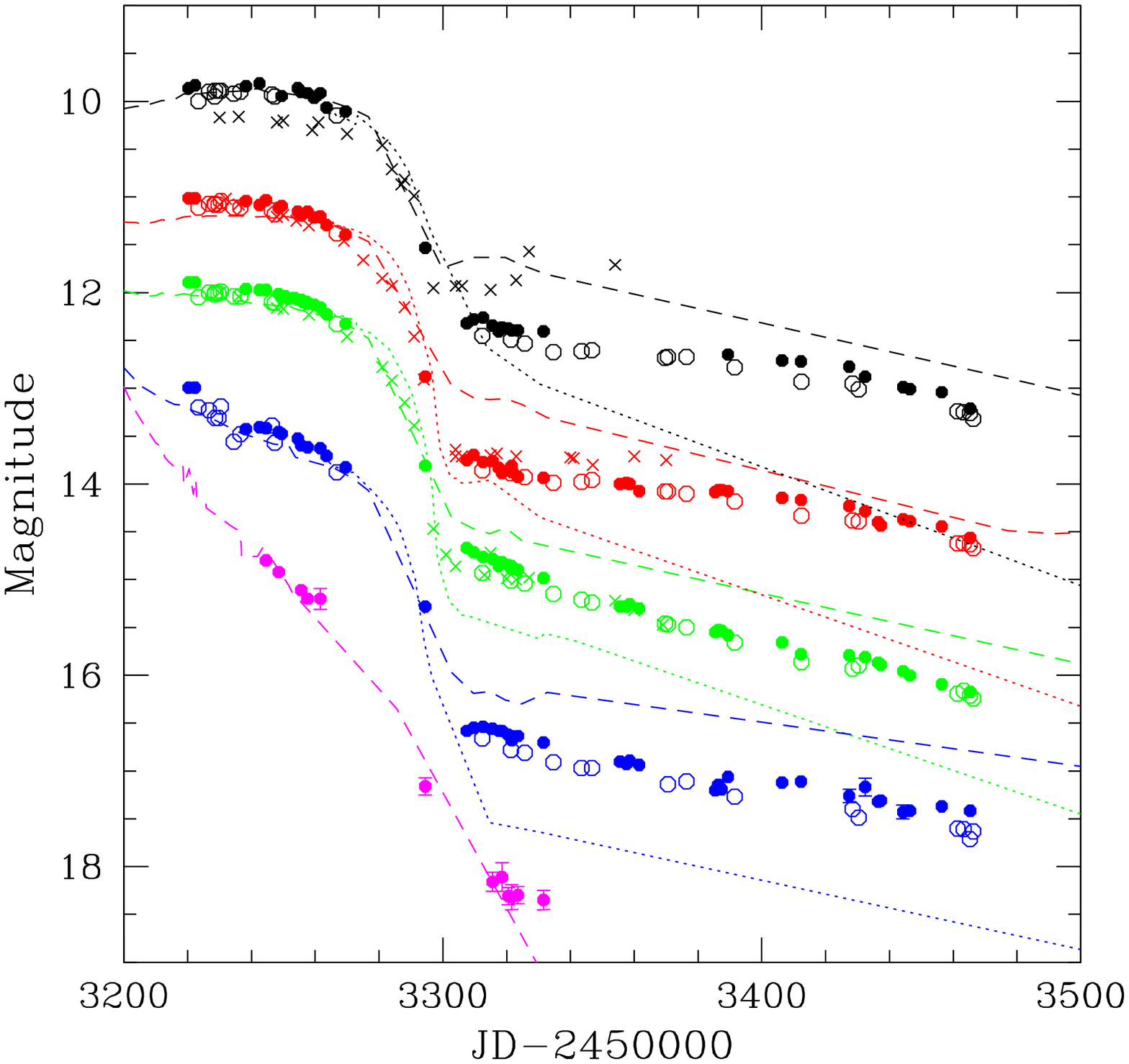}{$UBVRI$ light curves of SN 2004dj for the first 260 days
after discovery. The color coding and shifts of magnitudes 
are the same as on Fig. 2. Dots show our data, corrected for the 
luminosity of cluster S96, circles are for the magnitudes from Vink\'o et al. (2006),
and crosses show results from Zhang et al. (2006). Dashed lines are
the light curves of SN 1999em, and dotted lines show the light curves
of SN 2003gd}

Fig. 4 presents the comparison of our results with the data by
Vink\'o et al. (2006) and Zhang et al. (2006), and also the match
between the light curves of SN 2004dj and typical SNe II-P 1999em
(Leonard et al., 2002;
Elmhamdi et al., 2003; Hamuy et al., 2001) and 2003gd (Hendry et al., 2005).
The agreement of our data with the results by Vink\'o et al. (2006) is 
quite satisfactory, taking into acount the difference between the
calibrations of local standards and non-standard transmission of some of
our filters. The $VRI$ magnitudes by Zhang et al. (2006) were transformed to
standard system from photometry in their intermediate-band filters,
so large systematic differences can be expected, and we really see 
strong departures from our light curve in the $I$ band and in the $R$
at late stage. 
The comparison of light curves of type II-P SNe 2004dj, 1999em and
2003gd reveals the diversity of photometric evolution for the
objects of this class. We align the light curves in magnitudes so that 
they coincide at the plateau, and shift in time to match the early 
decline from the plateau. The differences are evident: SN 2003gd has
the largest drop from the plateau to the start of exponential tail and
so sign of flattening; for SN 1999em the decline in about one magnitude
less and the small flattening is evident. The light curves of SN 2004dj
lie between the curves for these two SNe, and the flattening in the $R$
and $I$ bands is the most pronounced.
 
\PZfig{12cm}{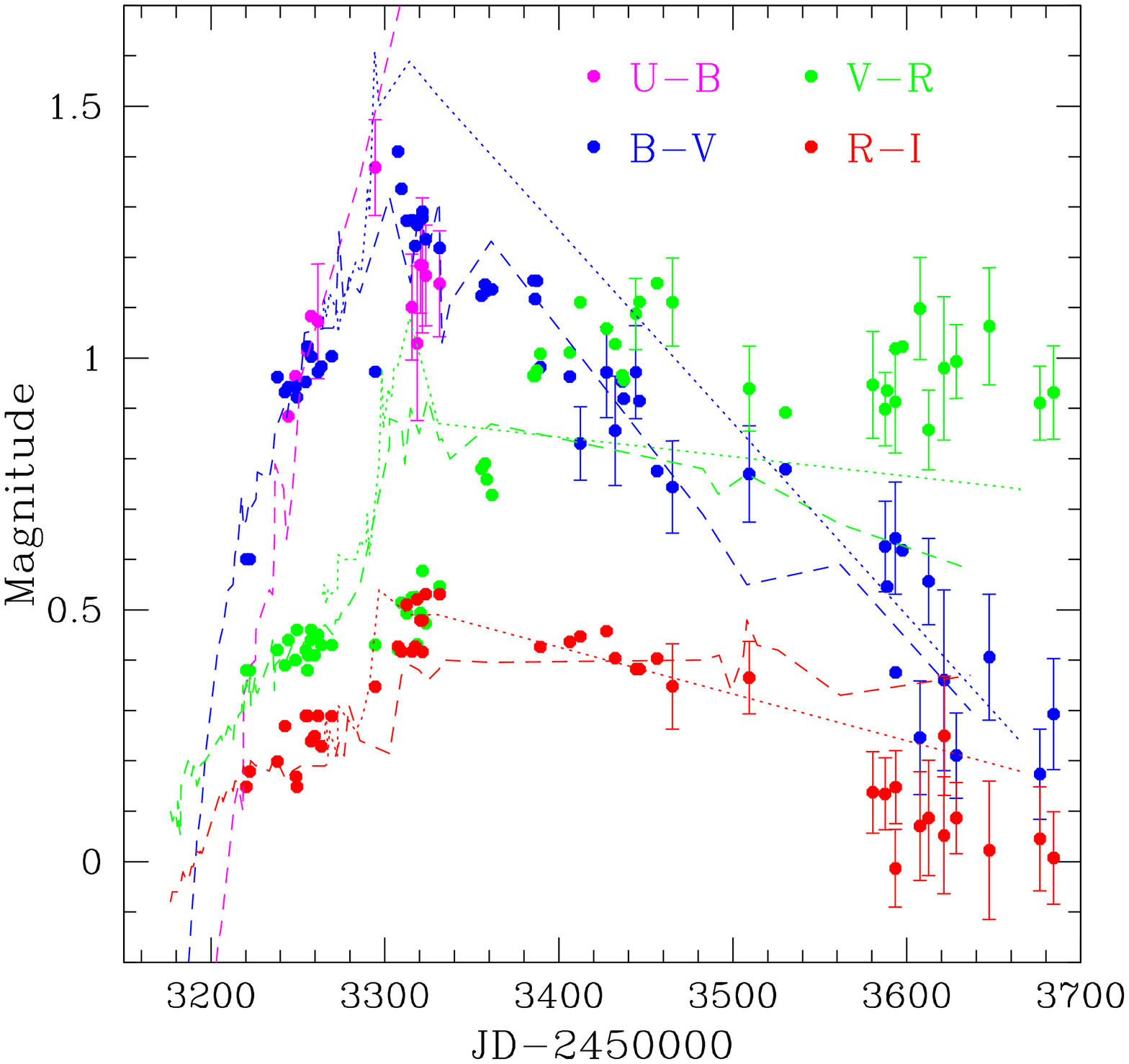}{The color curves for SN 2004dj compared with 
the curves for SN 1999em (dashed lines) and SN 2003gd (dotted lines),
shifted as reported in the text}

\PZfig{12cm}{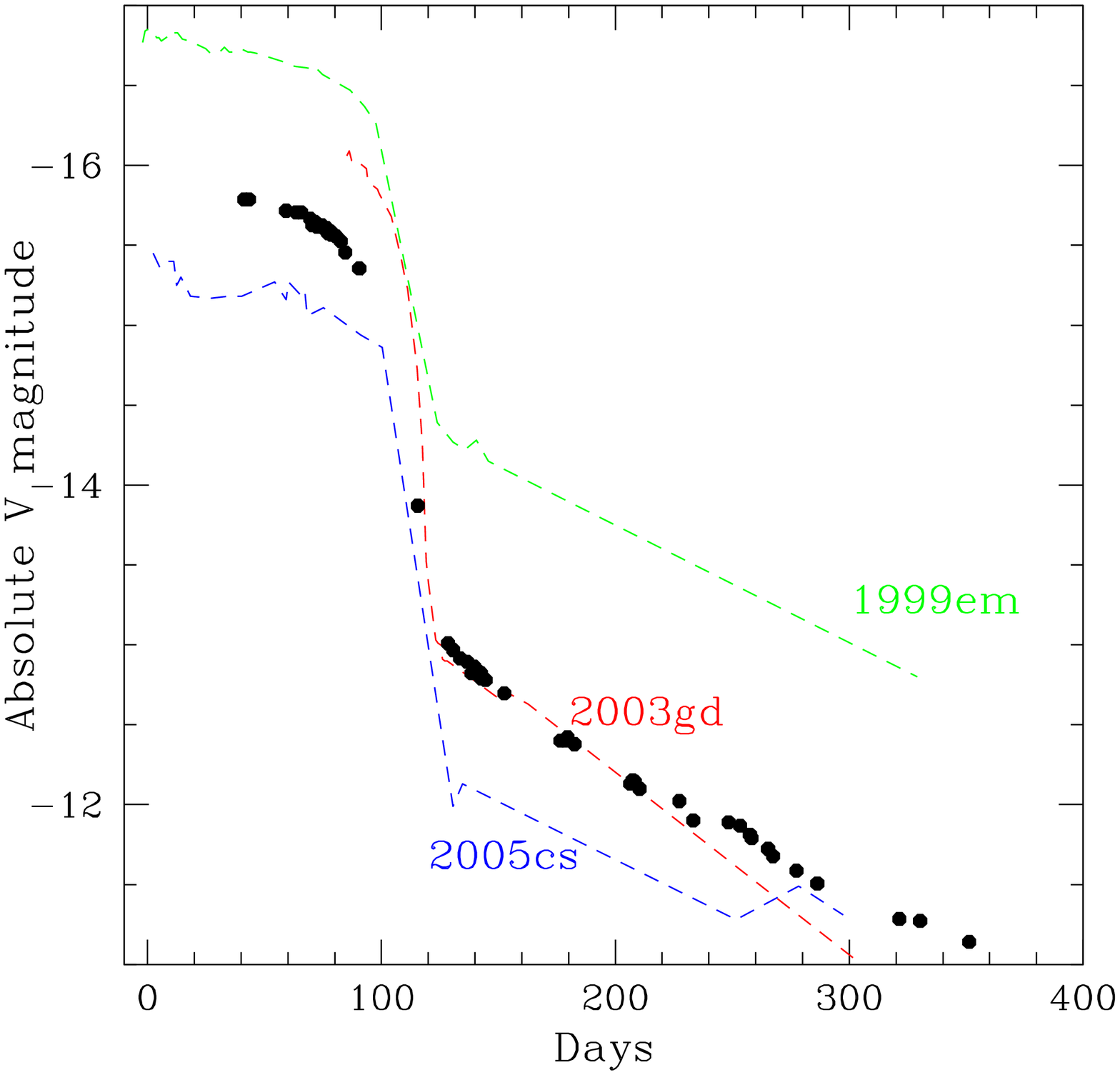}{The absolute $V$-band light curve of SN 2004dj
(black dots) compared to the light curves of SNe 1999em, 2003gd and 2005cs}

The color curves for the same objects are shown in Fig. 5.       
The color curves of SNe 1999em and 2003gd were shifted for better
alignment with the curves for SN 2004dj in the first week after discovery. 
The resulting shifts for SN 1999em in $U-B$, $B-V$, $V-R$ and
$R-I$ colors, respectively: -0.9, -0.4, -0.1, -0.1. For SN 2003gd in 
the same colors, except $U-B$, they are: -0.2, 0, -0.1. The data clearly
shows that SN 2004dj is bluer than SNe 1999em and 2003gd,
and the shape of the $V-R$ color curve is different 
after JD 2453280. As the total interstellar extinction for SNe 1999em and 
2003gd is quite small, with estimates of $E(B-V)$
in the range 0.075-0.1 for SN 1999em and 0.13-0.14 for SN 2003gd
(Elmhamdi et al., 2003; Hendry et al., 2005), our result suggests 
that extinction for SN 2004dj is also small, and perhaps the colors are
intrinsically bluer. We adopt for SN 2004dj $E(B-V)=0.07$, the value 
preferred by Vink\'o et al. (2006).    

The absolute $V$ band light curves of SNe II-P 2004dj, 1999em, 2003gd 
and 2005cs (Tsvetkov et al., 2006) are 
compared in Fig. 6. We adopted the following values of distance
and extinction: SN 2004dj: $D=3.1$ Mpc, $A_V=0.22$; SN 1999em:
$D=11.7$ Mpc, $A_V=0.3$; SN 2003gd: $D=9.3$ Mpc, $A_V=0.43$;
SN 2005cs: $D=8.3$ Mpc, $A_V=0.3$.
SN 2004dj is fainter at the plateau than the normal SNe II-P 1999em and
2003gd, but brighter than subluminous SN 2005cs. At the start of the tail
the luminosity of SN 2004dj is the same as for SN 2003gd. 

\bigskip
\bigskip
\PZsubtitle{The cluster S96 before and after the outburst of SN 2004dj}

The observations of SN 2004dj host cluster S96 attracted much attention,
because they can reveal the nature of SN precursor. The data 
published up to now were obtained before SN explosion, but we also
carried out photometry long after the outburst. The data in Table 3
report our PSF-photometry for S96 before explosion, and also 
at very late stage about 3.4 years past SN explosion. The results show
that the luminosity of the cluster in the $R$ band is the same before and
after explosion, but in the $B$ and $V$ bands it is slightly brighter
after outburst. If the brightness decline of SN 2004dj
continues at the rate we estimate, then at JD 2454420 it is about 
25 mag, and can add only about 0.002 mag to the luminosity of the cluster.
Of course, the use of PSF-photometry for non-stellar object may be
unjustified. 
We geometrically transformed the frames obtained at S100 on JD 2451929
(with K-585 CCD) and at C60 on JD 2454418 to a common pixel grid
defined by the images from S100 on JD 2454181 
and performed aperture photometry with
identical parameters. The results are presented in Table 4.           

{\renewcommand{\arraystretch}{0}
\begin{table}
\centering
\fn
\caption{Aperture photometry of cluster S96 before and after SN explosion}\vskip2mm
\begin{tabular}{cccccccl}
\hline
\rule{0pt}{4pt}&&&&&&& \\
JD 2450000+ & $B$ & $\sigma_B$ & $V$ & $\sigma_V$ & $R$ & $\sigma_R$ &
 Tel.\\
\rule{0pt}{4pt}&&&&&&& \\
\hline
\rule{0pt}{4pt}&&&&&&& \\
\strut 1929.48  & 18.41 & 0.07 & 17.95 & 0.05 & 17.66 & 0.04 & S100(K-585)\\
\strut 4181.44  & 18.05 & 0.04 & 17.86 & 0.04 & 17.64 & 0.05 & S100 \\
\strut 4418.52  & 18.29 & 0.04 & 17.87 & 0.03 & 17.76 & 0.04 & C60b2 \\
\hline
\end{tabular}
\end{table}
}

We can compare our results with the magnitudes of the cluster S96
before the explosion as estimated by Ma\'iz-Apell\'aniz et al. (2004),
Wang et al. (2005) and Vink\'o et al. (2006). Their data are, respectively:
$B=18.19; 18.29; 18.25; V=17.93; 18.07; 17.85; R=17.88; 17.52 $
(no $R$ band photometry is given by Ma\'iz-Apell\'aniz et al. (2004)).
The scatter is quite large, as can be expected for the photometry of
a non-stellar object superimposed on the bright background of the host
galaxy, and our data are in general agreement with these results. 

We may conclude that the brightness of the cluster is the same within the
errors of our magnitudes before and after the outburst. The only discordant 
estimate of $B$ on JD 2454181.44 is likely due to some accidental error, as 
PSF-photometry
on this image gives fainter magnitude than for the frame 
obtained on JD 2454418.52 at C60. 
We do not observe the decline of cluster luminosity which is expected 
after the explosion of a supergiant star, but the accuracy of our magnitudes 
is not sufficient to
detect the expected dimming by 0.02-0.1 mag (Ma\'iz-Apell\'aniz et al., 2004;
Wang et al., 2005).

\bigskip
\bigskip
\PZsubtitle{Conclusions}

We conclude that SN 2004dj is a normal SN II-P, but some 
peculiarities of photometric evolution are evident:
the flattening of the light curves in the $R$ and $I$ bands after
a drop from the plateau is more pronounced than for most of SNe II-P;
the shape of the color $V-R$ curve is different from that for typical
SNe II-P, and all colors may be systematically bluer.
The luminosity at the plateau $M_V \sim -16$ mag is quite normal.
As the luminosity at the tail is nearly the same for SN 2004dj and
SN 2003gd, we may assume that they produce similar amounts of $^{56}$Ni.
For SN 2003gd Hendry et al. (2005) obtained 
$M_{\rm Ni}=0.016 \pm 0.01$ $M_\odot$, and this is in good agreement with the
estimates for SN 2004dj from Chugai et al. (2005), Vink\'o et al. (2006)
and Zhang et al. (2006).  

\medskip

The work was partly supported by the Council for the Program of Support
for Leading Scientific Schools (projects NSh.433.2008.2, NSh.2977.2008.2).
 
\references

Chugai, N.N., Fabrika, S. N., Sholukhova, O. N., et al.,
2005, {\it Ast. L.}, {\bf 31}, 792

Chugai, N.N., 2006, {\it Ast. L.}, {\bf 32}, 739

Elmhamdi, A., Danziger, I.J., Chugai, N., et al., 2003,
{\it MNRAS}, {\bf 338}, 939 

Hamuy, M., Pinto, P.A., Maza, J., et al., 2001,
{\it Astrophys. J.}, {\bf 558}, 615 

Hendry, M.A., Smartt, S.J., Maund, J.R. et al., 2005, {\it MNRAS},
{\bf 359}, 906 

Korc\'akov\'a, D., Mikulasek, Z., Kawka, A.,
et al., 2005, {\it IBVS}, No. 5605

Kotak, R., Meikle, P., Van Dyk, S.D., et al., 2005,
{\it Astrophys. J.} {\bf 628}, L123

Leonard, D.C., Filippenko, A.V., Gates, E.L., et al., 2002,
{\it PASP}, {\bf 114}, 35 

Leonard, D.C., Filippenko, A.V., Ganeshalingam, M., et.al., 2006,
{\it Nature}, {\bf 440}, 505

Ma\'iz-Apell\'aniz, J.,
Bond, H.E., Siegel, M.H., et al., 2004, {\it Astrophys. J.},
{\bf 615}, L113

Nakano, S., 2004, {\it IAU Circ.}, No. 8377

Patat, F., Benetti, S., Pastorello, A., Filippenko, A.V., 2004,
{\it IAU Circ.}, No. 8378

Pooley, D., Lewin, W.H.G., 2004, {\it IAU Circ.}, No. 8390

Stockdale, C.J., Sramek, R.A., Weiler, K.W., et al., 2004,
{\it IAU Circ.}, No. 8379

Sugerman, B., Van Dyk, S.D., 2005, {\it IAU Circ.}, No. 8489

Tsvetkov, D.Yu., Volnova, A.A., Shulga, A.P., Korotkiy, S.A.,
Elmhamdi, A., Danziger, I.J., Ereshko, M.V., 2006,
{\it Astron. Astrophys.}, {\bf 460}, 769

Vink\'o, J., Tak\'ats, K., S\'arneczky, K., et al., 2006,
{\it MNRAS}, {\bf 369}, 1780

Wang, X., Yang, Y., Zhang, T., et al., 2005, {\it Astrophys. J.},
{\bf 626}, 89

Yamaoka, K., Ma\'iz-Apell\'aniz, J., Bond, H.E., Siegel, M.H., 2004,
{\it IAU Circ.}, No. 8385

Zhang, T., Wang, X., Li, W., et al., 2006,
{\it Astron. J.}, {\bf 131}, 2245

\endreferences
\end{document}